**Title:** Actomyosin contractility rotates the cell nucleus

**Author Affiliations:** Abhishek Kumar[1,+], Ananyo Maitra[2,+], Madhuresh Sumit[1], Sriram Ramaswamy[2,3,*] & G.V. Shivashankar[1,*]

[1]Mechanobiology Institute and Department of Biological Sciences, NUS, Singapore 117411
[2] Department of Physics, Indian Institute of Science, Bangalore 560012, India
[3]TIFR Centre for Interdisciplinary Sciences, Hyderabad 500075, India

[+]Equal contributors
*Corresponding authors

**Corresponding Authors:** G.V. Shivashankar, Mechanobiology Institute, National University of Singapore, T-Lab #05-01, 5A Engineering Drive 1, Singapore 117411.
Email: shiva.gvs@gmail.com
Sriram Ramaswamy, TIFR Centre for Interdisciplinary Sciences, 21 Brundavan Colony, Osman Sagar Road, Narsingi, Hyderabad, India 500075. Email:sriram@tifrh.res.in

**Abstract**

The nucleus of the eukaryotic cell functions amidst active cytoskeletal filaments, but its response to the stresses carried by these filaments is largely unexplored. We report here the results of studies of the rotational dynamics of the nuclei of single fibroblast cells, with the effects of cell migration suppressed by plating onto circular fibronectin-coated micro-fabricated patterns. The nucleus was observed to undergo noisy but coherent rotational motion. We show that this observation can be understood through a hydrodynamic approach in which the nucleus is treated as a highly viscous inclusion residing in a less viscous fluid of orientable filaments endowed with active stresses. Lowering actin contractility selectively by introducing blebbistatin at low concentrations drastically reduced the speed and coherence of the angular motion of the nucleus. Time-lapse imaging of actin revealed a correlated hydrodynamic flow around the nucleus, with profile and magnitude consistent with the results of our theoretical approach. Coherent intracellular flows and consequent nuclear rotation thus appear to be an intrinsic property of cells, possibly suppressed by specific mechanisms in some situations.

## Introduction

The nucleus is the largest and stiffest organelle in a eukaryotic cell (1). It is actively coupled to the dynamic cytoskeleton (2-5) by means of a variety of scaffold proteins: contractile (6) actomyosin complexes, microtubule filaments constantly undergoing dynamic reorganization, and load bearing intermediate filaments (2-4, 7-10). It has long been known that the nucleus both translates and rotates during cell migration (11-18). It is reasonable to suppose that such motions are a result of active processes in the cytoplasm, involving the cytoskeleton and molecular motors (2, 3, 7-10, 13, 16, 17). While a number of molecular players have been implicated in this context (2-5, 9, 10), the role of actomyosin contractility on nuclear dynamics has not been explored.

In this paper, we show that active stress is a critical component in determining nuclear movements. Fibroblast cells (NIH3T3) were plated on circular micro-patterned fibronectin surfaces to study the motion of the nucleus, independent of the effects of cell migration. Time-lapse imaging revealed a correlation between actin flow patterns and nuclear movement. We show that a hydrodynamic model of oriented filaments endowed with active contractile stresses (19-23) with the nucleus entering only as a passive inclusion, gives rise to the observed organized actin flow and nuclear rotation. The theoretical formulations in (24, 25) have predicted behaviors similar to ours. However, the context in which these works are set is different from ours, i.e., the dynamics of the cell nucleus is not the subject of these papers. In addition, the boundary conditions are different in detail. Reference (24) was in a Taylor-Couette geometry, i.e., there is no medium inside the inner circle, and Reference (25) was in a circular geometry without a central inclusion. Our observations suggest that nuclear rotation and circulating flows are an inherent property of the active cell interior under geometric confinement.

## Materials and Methods

### Cell Culture

NIH3T3 fibroblasts (ATCC) were cultured in low glucose DMEM (Invitrogen) supplemented with 10% Fetal Bovine Serum (FBS) (Gibco, Invitrogen) and 1% Penicillin-Streptomycin (Invitrogen). Cells were maintained at $37^0$C in incubator with 5% $CO_2$ in a humidified condition. Cells were trypsinized and seeded on fibronectin coated patterned surfaces for 3 hours before staining or imaging. For confocal imaging, untreated hydrophobic dishes (Ibidi) with the micropatterns were used. 65,000 cells were seeded on the micropatterned surfaces (with 10,000 patterns) for 30 minutes, after which the non-adhered cells were removed and fresh media was re-added in the dishes. All the control experiments were carried out without any treatment. Blebbistatin (Sigma) was serially diluted from high concentration (68mM) stock using filtered media to the final concentration (1.25μM).This minimizes the effect of any other solvent like DMSO. Nocodazole (Sigma), Latrunculin A (Sigma) and EHNA (sigma) were used at 0.25 μg/ml , 100nM and 20μg/ml respectively. All drugs were added after cells had fully adhered to the micropatterns and imaging was carried out atleast 1hour after the addition of the drugs.

### Preparation of PDMS Stamps and micro-contact printing

PDMS stamps were prepared from the PDMS Elastomer (SYLGARD 184, DOW Corning) and the ratio of curing agent to the precursor used was 1:10. The curing agent and the precursor were mixed homogeneously before pouring onto the micropatterned silicon wafer. The mixture was then degassed in the desiccator for at least 30 minutes to remove trapped air bubbles and cured at 80°C for 2 hours. Then, the stamps were peeled off from the silicon wafer. Micropatterned PDMS stamps were oxidized and sterilized under high power in Plasma Cleaner (Model PDC-002, Harrick Scientific Corp) for 4 minutes. 30μl of 100μg/ml fibronectin solution (prepared by mixing 27μl of 1xPBS to 1.5μg of 1mg/ml fibronectin and 1.5μl of Alexa 647 conjugated fibronectin) was allowed to adsorb onto the surface of each PDMS stamp under sterile condition for 20 minutes before drying by tissue. The PDMS stamp was then deposited onto the surface of an untreated hydrophobic dish (Ibidi) (for High-resolution imaging) to transfer the micro-features. Each patterned dish was inspected under fluorescent microscope to verify the smooth transfer of fibronectin micropatterns. To passivate the non-fibronectin coated regions of the dish, it was treated with 1ml of 2mg/ml Pluronic F-127 for 2 hours.

**Cell Transfection**
Transfection of Lifeact GFP (for imaging actin dynamics) and dynamitin (p50) subunit of dynactin complex (a gift from Alexander Bershadsky's lab, MBI, Singapore) in wt NIH3T3 cells was carried out using JetPRIMEpolyplus transfection kit. 1μg of plasmid was mixed in 100 μl of JetPRIME buffer by vortexing and spinning, 3.5 μl of JetPRIME reagent was then added and the mixture was again vortexed and spun. The mixture was incubated for 30 minutes and then added to a 50-60% confluent culture in 35mm dish. The cells were kept in the fresh media for 2 hours before adding the transfection mixture. Cells were incubated for 20 hours before plating them on the patterned substrate.

**Imaging**
Phase contrast imaging of cells on different geometrical patterns was done using Nikon Biostation IMq using 40x objective at $37^0$C in a humidified chamber with 5% $CO_2$. Confocal time lapse imaging of the cells transfected with Lifeact EGFP was carried out using Nikon A1R with 60x, 1.4 NA oil objective at $37^0$C in a humidified incubator with 5% $CO_2$.

**Image Analysis and quantifications**
Acquired images were processed and analysed using ImageJ software (http://rsbweb.nih.gov/ij/index.html). To determine the translational coordinates and the rotational angle of the nucleus, diagonally opposite nucleoli were manually tracked from the phase contrast image of the cell using the ImageJ plugin- MtrackJ (http://www.imagescience.org/meijering/software/mtrackj/). The nucleoli can be clearly seen in these images without any further fluorescence labelling (Fig. 1). The nucleoli were used to track the nuclear dynamics as their positions are stable during interphase of cell cycle (11, 26). All our analysis is confined to this time period during cell cycle. Nuclear translation is estimated through the instantaneous mean position of two nucleoli situated at roughly diametrically opposed points. Nuclear angle was determined from the slope of the line formed between the centroid and a nucleolus.
Particle image velocimetry (PIV) analysis was carried out using Matlab PIV toolbox-Matpiv between consecutive image frames separated by 1min. Images acquired were 512 X 512 pixels. Size of interrogation window was chosen to be 32 X 32 with an overlap of 50% between the consecutive time frames. "Single pass" method was used for calculating the velocities. The speed and angle of actin flow vectors (Fig. 3) was determined from the correlation of these interrogation windows with their neighbouring ones. Azimuthal ($v_\emptyset$) and radial ($v_r$) velocities were calculated from the actin flow speed. Velocity components in radial and angular coordinate are as follows: $v_r = \frac{1}{r}(x.v_x + y.v_y)$ and $v_\varphi = \frac{1}{r}(x.v_y - y.v_x)$, where, $r = \sqrt{x^2 + y^2}$ is the radial distance from the centre of nucleus and $v_x$ and $v_y$ are the actin flow velocity components determined from PIV analysis along x and y direction/axis measured from centre of nucleus. Quantifications were done using programs written in either LabVIEW 6.1 or MATLAB R2010a. All the graphs and curve fittings were carried out using OriginPro 8.1 (OriginLab Corporation, Northampton, USA).

**Results and Discussion**

**Geometric constraints on the cell affect the dynamics of the nucleus**

To assess nuclear dynamics independent of cell migration, we used micro-patterned fibronectin-coated substrates to confine cells to circular regions. The spread area of the cells (1600μm$^2$) was chosen to be close to the area that these cells occupy on unpatterned fibronectin coated dishes. Single cells were cultured on each patterned substrate and time lapse phase contrast imaging was carried out for about 8 hours. Fig. 1A shows phase contrast image and color-coded intensity profile image obtained by average intensity projection of the phase contrast time lapse images.

The translational and rotational movements of the nucleus were measured from the time lapse images. Fig. 1B displays typical trajectories of the nucleus. It is clearly seen that the nucleus both rotates and translates, as shown in Movie S1. Since XZ nuclear rotation is negligible, we resolve the dynamics into two-dimensional translation and rotation in the XY plane. Translation is estimated through the instantaneous mean position $(x_t, y_t) = \left(\frac{x_1+x_2}{2}, \frac{y_1+y_2}{2}\right)$ of two nucleoli situated roughly diametrically opposed points $(x_1, y_1)$ and $(x_2, y_2)$. The top inset to Fig. 1B shows a typical translation trajectory. Rotation is characterised by coordinates of one nucleolus relative to this mean. A typical rotational track for the nucleus is shown in bottom inset to Fig. 1B. Both translational and rotational movement of the nucleus during cell migration has been reported (11-17, 27) for many cell types including NIH3T3 which was used in all our experiments. For completeness, we document such motion here as well. Fig. S1A shows a representative DIC image of a monolayer of NIH3T3 cells cultured on glass bottom dishes. Time lapse images (Fig. S1B) of three cells from this field of view are presented with arrows showing the position of the nucleolus. Rotation and translation tracks of the nucleus are plotted for these cells in Fig. S1C and D respectively, showing large departures from its initial position and orientation. We also see that the nuclear rotation can be large even when the translation is small, for cells located near the middle of the monolayer and therefore constrained at their periphery. Thus, we suspect that migration is not the underlying cause of the observed rotation and confine cells by plating them onto fibronectin patterns of well-defined geometries to see whether this alone can produce nuclear rotation. Though here we present results only for the circular geometry, we note that nuclear rotation is a robust feature which is also observed on other shapes like squares and triangles (Fig S2). On rectangles with high aspect ratio, the rotation is suppressed presumably because of the narrower confinement and elongation of the nucleus due to the elongated cell shape (28).

To characterize the nuclear rotation, nuclear angle was determined from the slope of the line formed between the centroid and one of the nucleoli and is given by $\theta = tan^{-1}\left(\frac{y_r}{x_r}\right)$ where, $(x_r = x_1 - x_t)$ and $(y_r = y_1 - y_t)$. The instantaneous angular velocity was determined by $\frac{\delta\theta}{\delta t} = \frac{|\theta' - \theta''|}{\delta t}$, where, $\theta'$ and $\theta''$ are the nuclear rotational angle between consecutive time points with $\delta t = 5 min$. From this analysis, the mean instantaneous rotational velocity on circular patterns was found to be ~ $3.2^0$/min (Fig. S2). To explore the possible role of myosin induced contractility in these phenomena, we turn now to the active hydrodynamic theory (19-23) of the cell interior.

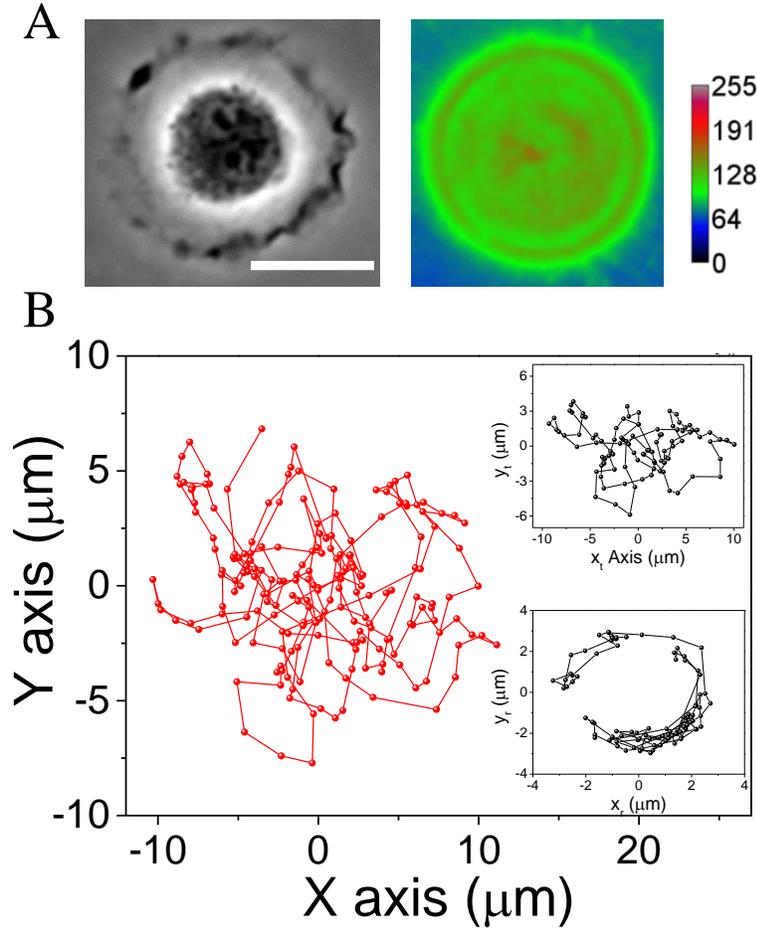

**Figure 1. Dynamics of the nucleus of a cell plated on a circular micropattern.** (A) Phase contrast image and corresponding color-coded average intensity projection of time lapse phase contrast image for cell plated on a circle of area 1600μm$^2$. (B) Typical XY trajectory of the nucleus. Inset: typical translational (top) and rotational (below) trace of nuclear motion.

**Active fluid with a central inclusion**

We show that cytoplasmic flows produced by actomyosin contractility are the minimal explanation for the observed rotation of the nucleus. To this end, we turn to the theoretical framework of active hydrodynamics (29-32). Contractile stresses carried by actomyosin, given an arrangement of filaments compatible with the cell shape imposed by the pads and the presence of the nucleus as an internal obstacle, lead to organized flows that rotate the nucleus. More detailed propulsive elements, e.g., pushing by microtubules anchored onto the nuclear surface (7, 27), while possibly present in the cell, are not a necessary part of the mechanism. Since the cell in the experiment is stretched, its height is smaller than its dimensions in the plane. We can therefore model it as a quasi-two-dimensional film with the hydrodynamics being cut off at a scale proportional to the height. We also assume an axisymmetric cell, and ignore treadmilling and the on-off kinetics of the motors. This highly simplified view of the cell still exhibits some key features of the dynamics found in the experiment.

We now present the equations of active hydrodynamics (29-32). The inner circular region represents the nucleus, which is taken to contain no active motor-filament complexes and is therefore modeled as a passive liquid drop of very high viscosity ($\eta_i$) - in effect undeformable. The outer annular region is the cytoplasm, which contains active orientable filaments. The inner fluid-fluid interface, i.e., the boundary between cytoplasm and nucleus, has tangential stress continuity and tangential velocity continuity, and the outer surface, the contact line of cell with pad, has no slip. We assume the filaments preferentially lie parallel to any surface with which they are in contact. In particular, they therefore lie tangent to both the inner and the outer boundaries.

The cytoplasmic medium is taken to consist of filaments suspended in the cytosol of viscosity $\eta \ll \eta_i$ (we will present results for $\eta/\eta_i = 1/5$). We assume the filaments are in a state of well-formed local orientation whose magnitude does not change so that it can be characterized completely by a unit vector or "director" field, $n(r)$, (33) at position $r$. Associated with the filaments is an active stress $Wn(r)n(r)$, where the parameter $W$ is a measure of actomyosin activity. The concentration of filaments and myosin is assumed uniform. Fluid flow in the cytoplasm is described by the hydrodynamic velocity field $v$. The equations of active hydrodynamics in steady state lead to a dynamic balance between shearing and relaxation of filaments (34),

$$\frac{1}{\Gamma} \delta_{ij}^T \frac{\delta F}{\delta n_j} = -v_j \partial_j n_j + \lambda_{ijk} \partial_k v_j \tag{1}$$

and force balance, ignoring inertia,

$$\nabla_j \sigma_{ij} = \zeta v_i \tag{2}$$

with total stress tensor

$$\sigma_{ij} = \frac{\eta}{2}[\nabla_i v_j + \nabla_j v_i] - P\delta_{ij} - Wn_i n_j + \lambda_{kij} \frac{\delta F}{\delta n_k} + \sigma_{ij}^0 \tag{3}$$

Here, $F = \int d^d x f$, $f = (K/2)(\nabla n)^2$ is the elastic free energy for the director $n$, with the Franks elastic constant $K$, $\delta_{ij}^T = \delta_{ij} - n_i n_j$, $\lambda_{ijk} = (1/2)(\delta_{ki}^T n_j - \delta_{kj}^T n_i) + (\lambda/2)(\delta_{ki}^T n_j + \delta_{kj}^T n_i)$ is a flow-orientation coupling and $\sigma^0 = (\nabla n) \partial f / \partial(\nabla n)$ is the Ericksen stress. We will work with a completely symmetric stress, $\sigma_{ij}^S$, built from $\sigma_{ij}$, which will give the same velocity field, due to angular momentum conservation (35, 36).

The coefficient $\zeta$ in (2) represents in a z-averaged sense the effects of confinement on the damping of velocities. It has two contributions: a viscous part $\eta/h^2$ arising because flows within an adhered cell of thickness $h$ in the vertical $z$ direction and no-slip at the base must in general have z-gradients on a scale $h$, and direct damping of flow through the kinetics of attachment and detachment of the cytoskeletal gel to the substrate. In our estimates below we retain only the viscous effect, so that $\zeta$ simply has the effect of screening the hydrodynamics at in-plane length-scales larger than $h$. Including attachment-detachment enhances $\zeta$. We use circular polar coordinates $r\varphi$ in the plane. Since we assume axisymmetry, the radial velocity vanishes because incompressibility implies $dv_r/dr + v_r/r = 0$, and $v_r = 0$ at both the interfaces. For force balance in the region corresponding to the nucleus we have to solve the equation

$$\eta_i \frac{d^2}{dr^2} v_\varphi + \frac{\eta_i}{r}\frac{d}{dr} v_\varphi - \frac{\eta_i}{r^2} v_\varphi = 2\zeta_i v_\varphi \tag{4}$$

where $\zeta_i = \eta_i/h^2$. The equation can be solved in terms of Bessel functions, with the constraint that $v_\varphi$ has to be 0 at $r = 0$. Continuity of tangential stress and velocity at the cytoplasm-nucleus interface gives the requisite number of boundary conditions.

Force balance in the azimuthal direction reads

$$\frac{d}{dr}\sigma^S_{\varphi r} + \frac{2\sigma^S_{\varphi r}}{r} = \zeta v_\varphi \tag{5}$$

Expressing the director $n = (n_\varphi, n_r) = (cos\theta, sin\theta)$ the steady state equation (1) for the orientation field reads

$$\frac{1}{\Gamma}\frac{\delta F}{\delta \theta} = \frac{v_\varphi}{r} + \lambda cos2\theta A_{r\varphi} + \Omega_{r\varphi} = (\lambda cos2\theta - 1)A_{r\varphi} \tag{6}$$

where $A$ and $\Omega$ are the symmetric and antisymmetric parts of the velocity gradient tensor.

Using (6), the $r\varphi$ component of (3) can be recast as a first order differential equation for $v_\varphi$.

$$\frac{d}{dr}v_\varphi = \frac{2W sin2\theta + 4\sigma^S_{\varphi r}}{2\eta + \Gamma(\lambda cos2\theta - 1)^2} + \frac{v_\varphi}{r} \tag{7}$$

Thus, we have two first order equations, (7) and (5), and one second order equation (6) to solve, which we solve numerically.

A noticeable and robust feature of the solution (inset to Fig. 3A) is the presence of a maximum in the magnitude of the velocity some distance from the nucleus. This results from a combination of vanishing velocity at the outer boundary and the nuclear centre, and continuity of velocity and shear stress at the fluid-fluid interface. Note that our description does not include chiral effects, so that equivalent solutions with either sense of rotation are obtained.

The competition between active stresses that promote flow and orientational relaxation, that inhibits it, is contained in the dimensionless combination $\alpha = W/\zeta D = W h^2/K$ (32, 37). Accurate estimates of parameters for our system are not easy to make. The cytoskeletal active stress, $W$, is generally argued to be in the range 50-1000 Pa (38). Frank constants for actin nematics appear to be 2-20 pN (39), as in ordinary thermotropic nematics. The thickness $h$ of the spread cell in our experiments is about 1/5 of the lateral extent. For a spread cell area of 1600μm² we therefore estimate $h = 8 μm$. Taken together, this leads to $α \sim 200$. However, if attachment-detachment contributions to $ζ$ are included, $α$ will be lowered substantially. From the active hydrodynamic model we know (31) that the system is quiescent for small values of this parameter. However, $α = 4.9$, for which we present the results, is already sufficient to produce a spontaneous flow. Increasing $α$ leads to increasingly complicated flows which we have only begun to explore. We do not attempt a detailed comparison between the observed and the theoretical flow patterns. However, the conclusion about the maximum of the velocity being away from the nucleus rests purely upon the confining geometry, and we expect that the time and angle averaged velocity profile, measured from the experiment, will have a peak away from the nuclear boundary.

For $α \gg 1$ i.e an unbounded, oriented active fluid, one expects (31) spontaneous velocity gradients of order $\frac{W}{η}$. In (32) and (37) it was shown that the presence of confinement on a scale $h$ modifies the above conclusion giving a characteristic rate $\frac{W}{η}F\left(\frac{h}{l}\right)$ (37) where $F(x) \to 1$ as $x \to \infty$ and $x^2$ for $x \to 0$, where $l$ is the in-plane scale associated with observation. In our case, $\frac{h}{l}$ is 1/5. Thus, the rotation rate should be of the order of $0.1 W/η$. Using the arguments of (38) this estimate turns out to be of the order of a few degrees/min. This is reassuringly consistent with the magnitude obtained from the experiment.

The two predictions we can make based on this simple model are that actomyosin is crucial for nuclear rotation, and that the angle and time averaged angular velocity will be maximum away from the nucleus. We perform a series of experiments to check these. In the next sections, we study the contribution of actomyosin contractility, a critical cytoplasmic regulator of nuclear prestress (40, 41), to the translational and rotational dynamics of the nucleus.

**Role of actomyosin contractility on nuclear dynamics**

We test the role of contractility on nuclear dynamics to validate the theoretical predictions based on active fluids with an inclusion. Actomyosin contractility was altered by treating cells fully spread on the circular patterns with low concentration of blebbistatin, an inhibitor of the myosin II motor. On treatment with blebbistatin, the instantaneous angular velocity significantly decreases to $1.7^0$/min when compared to control patterns $3.2^0$/min. Live cell fluorescence confocal imaging was carried out to simultaneously visualize actin flow dynamics and nuclear rotation. Cells were transfected with lifeact-GFP to label actin in live condition (Fig. 2). Time lapse confocal imaging of actin revealed a retrograde flow and its remodeling around the nucleus (Fig. 2A). To quantify the flow pattern (Movie S2), we carried out particle image velocimetry (PIV) analysis using MatPIV. This revealed flow vectors tangential to the nuclear boundary with direction and magnitude correlated with that of the nuclear rotation as shown in Fig. 2A and Movie S3. Velocity field maps of actin flow were determined in small regions throughout the cell (Fig. 2A, last panel and Movie S4). A circulating flow, required to rotate the nucleus, is clearly seen (Fig. 2A, middle panel and Movie S3 and S4). Interestingly, upon blebbistatin treatment, inward flow of actin (Movie S5-S7), presumably driven by treadmilling, was not significantly affected. However, the azimuthal speed can be seen (Fig. 2B) to decrease substantially, despite some scatter in the data. The circulation of flow around the nucleus was lost concurrent with the loss of nuclear rotation (Fig. 2B, middle panel and Movie S6).

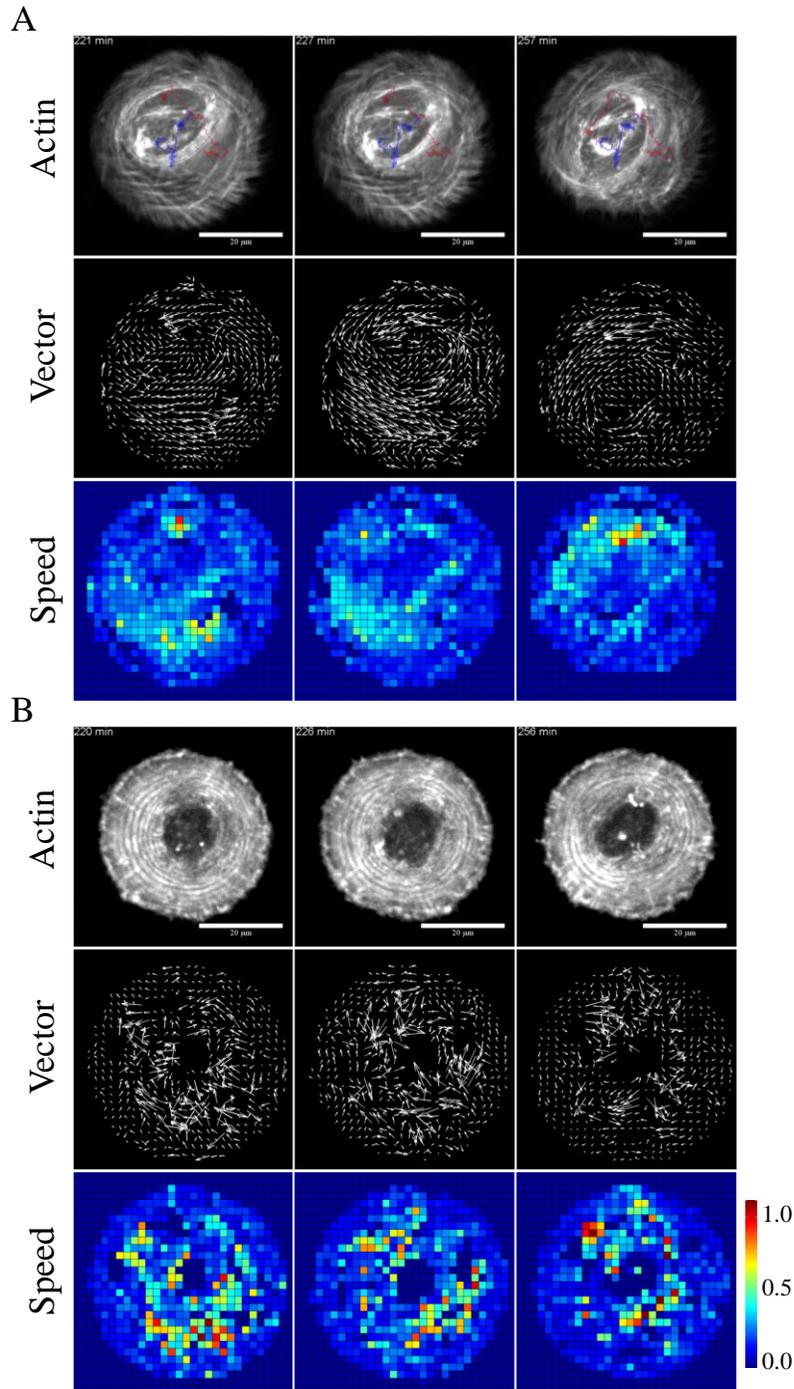

**Figure 2. Visualization of actin flow patterns during nuclear rotation in both control and blebbistatin treated cells plated on circular geometry.** (A) Top panel: Tracks of two nucleoli (red and blue) showing both translational and rotational dynamics. Scale bar = 20μm. Corresponding actin flow vectors (middle panel) and speed (last panel) was determined by particle image velocimetry (PIV) analysis using MatPIV for control (A) and blebbistatin (B) treated cells. The flow vectors have been scaled to 3 times for better visibility. Color code: 0.0 - 1.03μm/min.

Further, we plot in Fig. 3A and B, the angle averaged azimuthal velocity $v_\varphi$, with and without blebbistatin respectively, inferred from PIV as a function of radial distance from the centre of the nucleus. For comparison we also show the radial velocity, $v_r$ (Fig. 3C and D). Note that the graphs start from the edge of the nucleus. As predicted from the theory (inset to Fig.3A), the azimuthal velocity peaks away from the nuclear boundary in the control cells. In blebbistatin treated cells, by contrast, the velocity is 0, leading to the loss of nuclear rotation. However, $v_r$ is small in both cases, albeit with slightly larger fluctuations in the presence of blebbistatin.

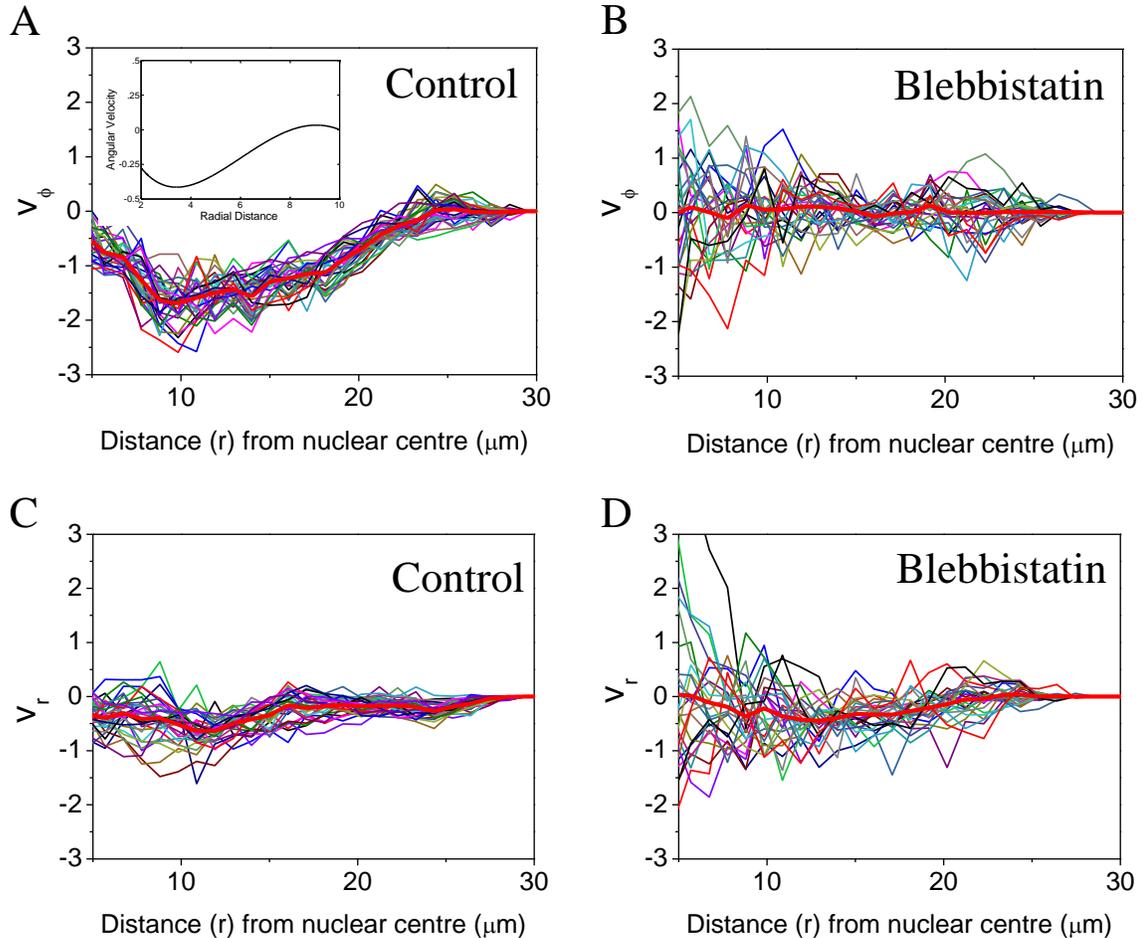

**Figure 3. Azimuthal and radial velocity of actin flow for cells plated on circular geometry.** Plot of $v_\varphi$ and $v_r$ from velocity vectors of actin flow for control (A) and (C); and blebbistatin treated cells (B) and (D). Each color represents single time point for cells. Thick red curve is mean of various such realizations (N=30 for control and N=25 for blebbistatin treated cells). Inset to 3A: a typical angular velocity vs. radial distance curve obtained by solving the equations (5)-(7).

Since there have been suggestions in the literature that nuclear rotation is primarily driven by microtubules, we also visualized the nuclear movement in cells (on triangular patterns) treated with nocodazole, which interferes with microtubule polymerization. Nuclear rotation is still seen, but is irregular and not consistent in direction (Movie S8). This is consistent with the idea that actomyosin contractility continues to drive flow, but the nocodazole-induced loss of the microtubule cage normally found around the nucleus delocalizes the nucleus thus making the rotation incoherent. We also checked that suppressing dynein activity, by the addition of EHNA and by over-expressing dynamitin (p50 sub-unit of dynactin), does not have a significant effect on the reported rotation. Thus, we can be reasonably confident that dynein-microtubule activity is not the major cause of nuclear rotation for cells on micropatterns. However, the detailed effect of microtubules on nuclear rotation remains to be studied, and a theoretical study of a liquid-drop nucleus free to move around among active filaments remains to be done.

Further, we disrupted the actin filaments using latrunculin A. However, the cell did not adhere properly to the geometric pad due to the loss of actin filaments, and the motion of the nucleus became noisy and random (Movie S9). We conjecture that the higher instantaneous nuclear rotational velocity upon latrunculin B treatment reported in moving cells by Gerashchenko et al. (42) results from a lowering of cytosol viscosity due to the disintegration of the actin filaments. Since the flow is not coherent, we did not observe a steadily rotating nucleus in our experiments.

**Conclusions**

While a number of components including cytoskeleton and motor proteins have been implicated to drive nuclear rotation in migrating cells, our results on constrained single cells suggest a central role for actomyosin contractility. We offer a simple theoretical explanation for the rotation in which the nucleus is modeled as a nearly rigid inclusion in the cytoplasm treated as a fluid containing filaments endowed with intrinsic stresses. The result is an angular velocity profile with a maximum magnitude at a radial position intermediate between the nucleus and the cell periphery, as observed in the experiments, and nonzero at the nuclear surface, corresponding to nuclear rotation. The predicted magnitude of the rotation rate based on plausible estimates of material parameters is also consistent with the measurements. That blebbistatin treatment greatly suppresses the flow lends support to our proposed generic mechanism. This does not imply that the nucleus of every cell-type will rotate in all situations. At least two mechanisms could contribute to suppressing the generic instability that leads to circulating flows. One, in the absence of a reasonably rigid geometry, the cell boundary is free to change shape. This would disrupt the imposed boundary orientation of the filaments, and hence the orderly pattern of active particles needed to drive a coherent flow. Two, the apical actin fibers (40, 41, 43, 44), absent in square and circular geometries, present to some extent in triangular geometries, and very well formed in elongated geometries, bear down on and thus enhance the friction on the nucleus, suppressing its motion.

Collectively, our results highlight the importance of actomyosin contractility for nuclear dynamics. We hope our work leads to a search for nuclear rotation in a wider range of systems and settings, whether such rotation has biologically significant consequences, and a deeper understanding of how the cells suppress such effects.

**Acknowledgements**

A.K., M.S., and G.V.S. thank the Mechanobiology Institute (MBI) at the National University of Singapore (NUS) for funding and MBI facility. A.M. thanks TCIS, TIFR Hyderabad for support and hospitality, and S.R. acknowledges a J.C. Bose fellowship.

# SUPPORTING MATERIAL

**Title:** Actomyosin contractility rotates the cell nucleus

**Author Affiliations:** Abhishek Kumar[1,+], Ananyo Maitra[2,+], Madhuresh Sumit[1], Sriram Ramaswamy[2,3,*] & G.V. Shivashankar[1,*]

[1]Mechanobiology Institute and Department of Biological Sciences, NUS, Singapore 117411
[2] Department of Physics, Indian Institute of Science, Bangalore 560012, India
[3]TIFR Centre for Interdisciplinary Sciences, Hyderabad 500075, India

[+]Equal contributors
[*]Corresponding authors

**Corresponding Authors:** G.V. Shivashankar, Mechanobiology Institute, National University of Singapore, T-Lab #05-01, 5A Engineering Drive 1, Singapore 117411.
Email: shiva.gvs@gmail.com
Sriram Ramaswamy, TIFR Centre for Interdisciplinary Sciences, 21 Brundavan Colony, Osman Sagar Road, Narsingi, Hyderabad, India 500075. Email: sriram@tifrh.res.in

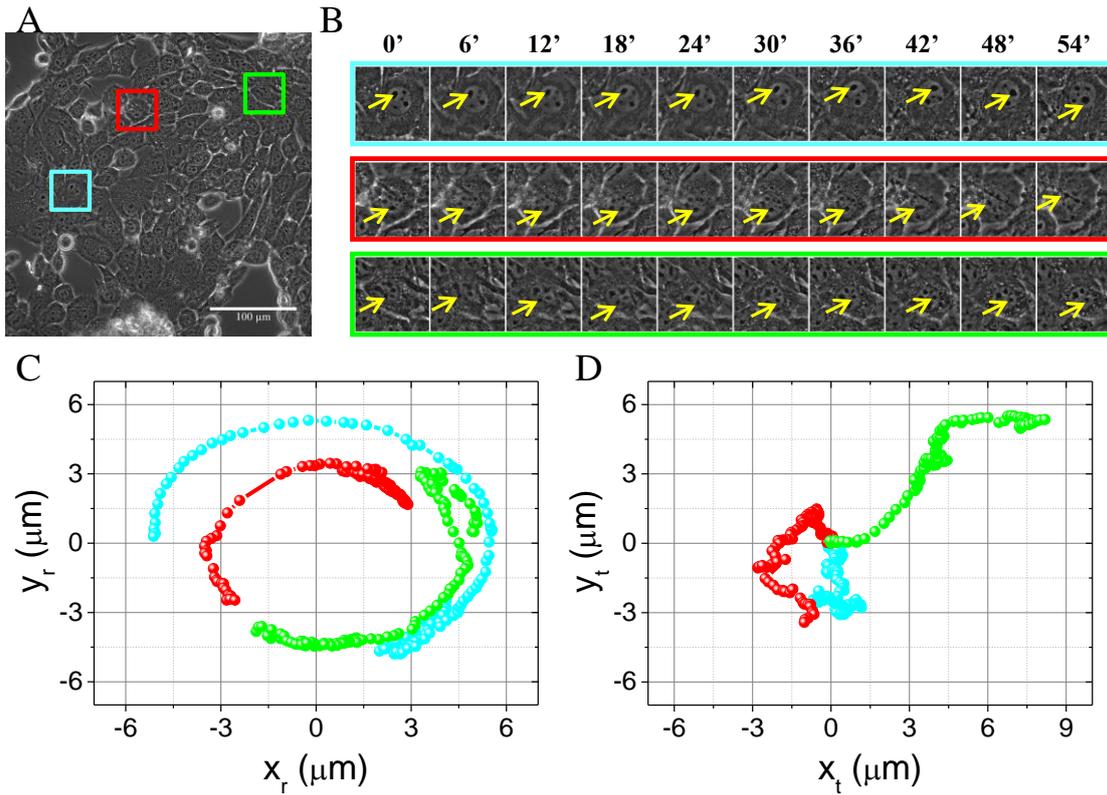

**Fig. S1.** Nuclear rotation in a monolayer of NIH3T3 cells. (A) Representative DIC images of NIH3T3 cells growing as a monolayer on glass bottom culture dishes. Scalebar = 100µm. (B) Time lapse images of single cells from different regions (cyan, red and green rectangles shown in (A)) of the monolayer. Yellow arrows show the position of one of the nucleoli in the time series. Time points are indicated at the top of each image. Plot of rotational (C) and translational (D) movement of the nucleus for the three cells shown in (B).

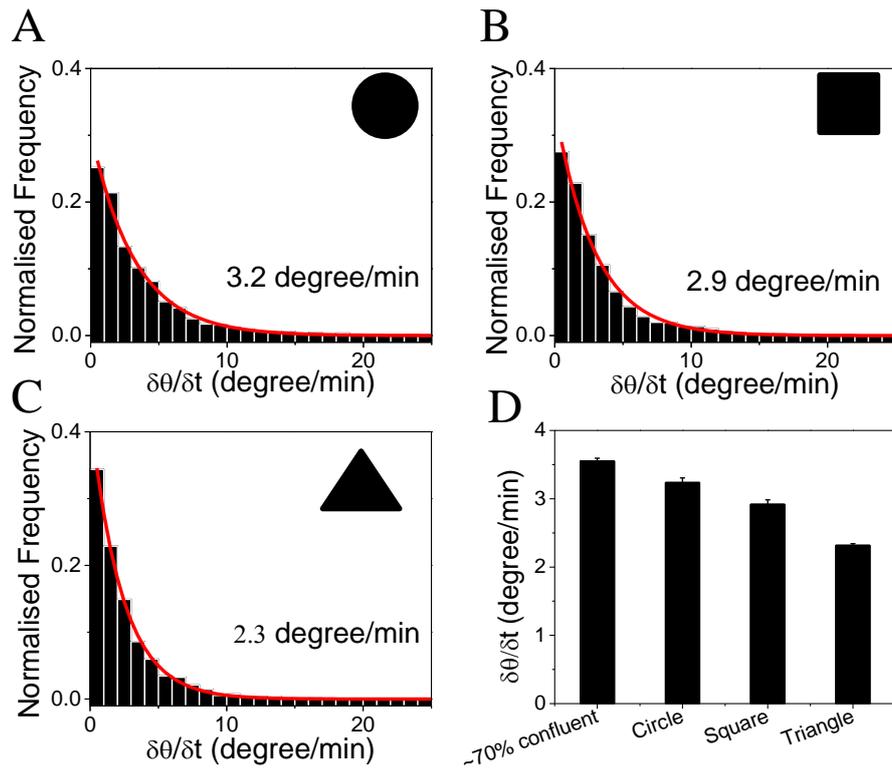

**Fig. S2.** (A-C) Histograms of instantaneous (absolute) angular velocity of the nucleus (vertical box) and fit to exponential distribution (red line) for cells on circular, square and triangular patterns respectively. Mean of the instantaneous angular velocities for each pattern is quoted in the corresponding plot. (D) Comparison of mean of instantaneous angular velocities of the nucleus for cells plated on various patterns and in 70% confluent culture. Error bars are SEM.

**Supplementary Movies Legends**

**Movie S1.** Phase contrast images of cell on circular geometry. Images were acquired every 5 min. Scale bar = 20µm.

**Movie S2.** Actin flow pattern on circular geometry.

**Movie S3.** Actin flow vectors on circular geometry.

**Movie S4.** Actin velocity field maps on circular geometry.

**Movie S5.** Actin flow pattern in cells treated with blebbistatin on circular geometry.

**Movie S6.** Actin flow vectors in cells treated with blebbistatin on circular geometry.

**Movie S7.** Actin velocity field maps in cells treated with blebbistatin on circular geometry.

**Movie S8.** Phase contrast images of a cell on triangular geometry treated with 0.25µg/ml Nocodazole after the cell fully adhered to the pattern. The image for the first frame was acquired 1 hour after the cell was treated with the drug. Images were then acquired every 5 min. Scale bar = 20µm.

**Movie S9.** Phase contrast images of a cell on triangular geometry treated with 100nM Latrunculin A after the cell fully adhered to the pattern. The image for the first frame was acquired 1 hour after the cell was treated with the drug. Images were then acquired every 5 min. Scale bar = 20µm.